\documentclass[aps,preprint,groupedaddress]{revtex4}
\usepackage{epsfig,amsbsy,bm}

\newcommand{\eref}[1] {(\ref{#1})}
\newcommand{\Eref}[1] {Eq.~(\ref{#1})}
\newcommand{\Fref}[1] {Figure \ref{#1}}

\begin{document}
\bibliographystyle{apsrev}

\baselineskip = 8mm\title
{Tailoring the waveforms to extend the high-order harmonic
generation cut-off }

\author{I. A. Ivanov\footnote[1]{Corresponding author:
Igor.Ivanov@.anu.edu.au}
and A. S. Kheifets}
\affiliation
{Research School of Physical Sciences and Engineering,
The Australian National University,
Canberra ACT 0200, Australia}\date{\today}

\begin{abstract}
Increase of the cut-off value in the high order harmonics
generation process is demonstrated for a special case of the
driving field composed of several harmonics of a given
frequency. It is shown  that a moderate, of the order of 20\%,
increase in the cut-off value can be achieved. This result
possibly
constitutes an upper limit for the increase in the cut-off value,
attainable for a class of the waveforms considered in the paper.
\end{abstract}


\maketitle

\section{Introduction}

High order harmonic generation (HHG) is a nonlinear atomic
process which can be described using a simple classical
picture \cite{Co94,hhg1,hb93}. Driven by a strong
electromagnetic (EM) field, the atomic electron emerges into
the continuum with zero velocity at some particular moment
of time. At a later time, the classical electron trajectory
returns to the nucleus, where the electron can recombine and
emit a photon. The frequency of the emitted photon is
determined by the amount of energy acquired by the electron
and the atomic ionization potential (assuming that the
electron recombines to the ground state).
The classical analysis shows that, for the monochromatic EM field
of the form $F_0\cos{\Omega t}$, 
the kinetic energy
of the electron returning back to the nucleus
cannot exceed the value of $3.17 U_{\rm p}$, 
where $\displaystyle U_{\rm p}=F_0^2/(4\Omega^2)$
is the ponderomotive potential. 
This leads to the
well-known $I_{\rm p}+3.17 U_{\rm p}$  cut-off rule for the
maximum harmonic order. Here $I_{\rm p}$ 
is the atomic ionization potential.

The quantum counterpart of the classical model \cite{hhgd}
assumes, that the released electron moves only under the
action of the EM field neglecting the influence of the
atomic potential ( the so-called strong-field approximation
- SFA \cite{Keldysh64,hhgd}.  
The SFA employs the analytical Volkov states, which
makes the problem tractable. The classical returning
trajectories emerge  as extrema in the saddle-point
analysis of the quantum-mechanical amplitudes computed
within the SFA \cite{hhgd}.

The aforementioned classical and quantum-mechanical results
correspond to the pure cosine form of the driving EM
field. Available pulse-shaping techniques \cite{pshape} make
it possible to modify the HHG characteristics by suitably
tailoring the driving EM field. This problem belongs to a
rapidly developing field of the quantum optimal control
\cite{tutorial}.

Several aspects of the optimal control of the HHG 
process were addressed in the literature. In the paper
\cite{wavelet1}, the emission intensity of a given harmonic order
was optimized by tailoring the laser pulse.  In
Ref.~\cite{opt2}, the emphasis was placed on optimizing the
particular high order harmonics from which single attosecond
pulses could be synthesized. Both these works employed the
so-called genetic algorithm, which mimicked the natural
selection process by introducing the mutation procedure and
suitable fitness function emphasizing the desired properties
of the target state. Numerically, this procedure
requires multiple solution of the time-dependent
Schr\"odinger equation (TDSE) which may constitute a
considerable computational task if one is interested in
formation of HHG in real atomic systems.

If the desired goal is to increase the harmonics cut-off
order, there is a possibility to find the optimum field
parameters using a purely classical approach based on the
electron trajectory analysis \cite{kinsler,hhgmcol2}.
In the paper by \citet{kinsler} such an analysis, supplemented by
the quantum calculation relying on the genetic algorithm,
was used to show, that the optimum waveform allowing to
maximize the recollision energy is a linear ramp with
the DC offset $\displaystyle F(t)=\alpha t+\beta$
for $t\in (0,T)$, $T$ being the period of oscillations.
Such a form has been shown to provide an
absolute maximum of the kinetic energy of the electron at
the moment of its return to the nucleus. This energy was
approximately 3 times larger than the corresponding energy
for the pure cosine wave with the same period and field
intensity \cite{hhgmcol2}. To
avoid using a strong DC field in practice, it was suggested in
\cite{hhgmcol2}, that it could be replaced by an  AC field
of a lesser $\Omega/2$ frequency, while the linear ramp could
be replaced by a combination of the harmonics with
frequencies $n\Omega$. Here $\Omega$ is the frequency
corresponding to the oscillation period $T$, $n$ is integer.
The overall pulse has thus a period of $2T$, rather than
$T$.  The weights corresponding to different harmonics
constituting the pulse were found by means of the genetic
algorithm.  Results of the quantum calculation relying on
SFA reported in \cite{hhgmcol2} confirmed, that this
waveform allowed to achieve considerable increase in the
cut-off position.

In the present work, we address a related question: what gain
in the HHG cut-off can be achieved if we use the driving EM
field with the waveform composed of the harmonics with the
multiple frequencies $n\Omega$. In other word, we demand
the driving EM pulse to be strictly $T$-periodic and such, that
its integral over a period is zero (i.e., no DC component is
present). It turns
out, that a moderate increase in the cut-off position is
possible in this case. A simpler case of adding the second
harmonic $2\Omega$ to the waveform was considered in earlier
works \cite{mauritsson:013001,zeng:203901}. 

We supplement the classical trajectory analysis by a
quantum mechanical TDSE calculation of the HHG process in
the lithium atom. 
Choice of this particular target was motivated by the 
experiments on the
laser field ionization of magneto-optically trapped (MOT) Li
atoms \cite{Steinmann07}. 
Numerical solution of TDSE takes
full account of the effect of the atomic potential. Such
a calculation ensures, that the effect of the
extended HHG cut-off, which we report below,
is not an artifact of a simplified
treatment. 

We shall consider below EM fields for which the field amplitude
$F(t)$ is a periodic function of time, having a
fixed period $T$. The field intensity for such EM fields can be expressed
as $\displaystyle W={c\over 4\pi T}\int\limits_0^T F(t)^2\ dt$,
where $c$ is the speed of light.
For
the monochromatic EM field $F(t)=F_0 \cos{\Omega t}$ this gives
the well-known relation
$\displaystyle W={F_0^2c\over 8\pi}$.
Throughout the paper we shall use the
atomic units. The unit of the EM field intensity 
corresponding to the 
unit field strength $F_0 = 1~{\rm a.u.} = 5\times 10^9$ V cm$^{-1}$
is $3.51\times 10^{16}$ W cm$^{-2}$ \cite{shreview}.
The field intensity of the 
monochromatic wave $F_0 \cos{\Omega t}$ can thus be expressed as
$W=3.5\times 10^{16} F_0^2$, if
field intensity is measured in W/cm$^2$ and the field
strength is expressed in the atomic units. From the expressions
above it is clear, that $T$-periodic EM fields with different $F(t)$ but 
equal values of  $\int\limits_0^T F(t)^2\ dt$,
will have the same intensities. In particular,
EM field $F(t)$ will have the same intensity as the monochromatic
wave $F_0 \cos{\Omega t}$ of the same period
if $\int\limits_0^T F(t)^2\ dt=T F_0^2/2$.

\section{Theory}

\subsection{Classical approach}

We begin with a purely classical problem of finding returning
trajectories of an electron moving in a periodic EM field
with a given period $T$, corresponding frequency $\Omega=2\pi/T$,
and which does not contain a DC component:
\begin{equation}
F(t)=2{\rm Re} \sum\limits_{k=1}^{K} a_k e^{i k \Omega t}
\label{comp}
\end{equation}

The field is assumed to be linearly polarized along the $z$-axis.
Our task is to find the set of coefficients $a_k $ in
\Eref{comp} for which electron returning to the nucleus
possesses the highest possible kinetic energy. For this
problem to be well-defined, we must impose some restrictions
on the possible choice of this set. A natural requirement is
that only the fields $F(t)$ 
of the same intensity
are to be
considered. This implies that $ 4\sum\limits_{k=1}^{K}
|a_k|^2=F_0^2$, where $F_0$ is amplitude of the monochromatic
waveform $F_0 \cos{\Omega t}$ having the same intensity.

As it is customarily done in the classical 3-step model of
HHG, we neglect the influence of the atomic core on the
electron motion.  We solve the classical equations of motion
of electron in the EM field
with the initial conditions $z(t_0)$=0, 
$\dot z(t_0)=0$. Here $t_0$ is the moment of time when the atomic
ionization event occurs.
In the classical calculation,
we do not introduce any envelope
function to describe the EM field, i.e., as a driving
force in the classical equations of motion,
we use the flat envelope pulse of infinite duration. 
This is permissible, since
in the quantum calculation
presented in the next section we shall use 
a pulse long enough, so that all transient effect,
as well as all 
effects due to the finite duration of the pulse (such as
dependence on the carrier phase) become unimportant.
The results of both calculations can, therefore, be legitimately
compared, and, as we shall see,
will give qualitatively similar results.

We are interested only in the
returning trajectories for which $z(t_1)$=0 for some
$t_1$. For such trajectories, we compute the kinetic energy
$E$ at the moment of return. 

We use the following field parameters: $I=10^{12}$ W/cm$^2$,
$F_0=0.0053$~a.u., $\Omega=0.185$ eV (6.705 $\mu$m). In this and the
subsequent section we consider the case of the Li atom with
the ionization potential $I_p=0.196$ a.u. For this set of the
field and atomic parameters, the value of the Keldysh
parameter $\gamma=\sqrt{I_p/2U_p}$=0.8. 

Our choice of the field
parameters was motivated, primarily, by the following reason.
We need to choose a combination of $F_0$,  $\Omega$, and 
$I_p$ such that the picture of the HHG process \cite{hhgd},
which establishes the connection of HHG with returning classical
trajectories remained valid. Among the conditions of the validity of
this picture are the requirements, that depletion of the ground state
can be ignored, and that the value of the
Keldysh parameter should be less than one \cite{hhgd}.
For the lithium atom, with its small ionization
potential, we have a rather narrow corridor of the field parameters,
which satisfy both these requirements.
For the field parameters thus defined we have the 
value $F_0/\Omega^2\approx 115$ a.u. for the excursion radius
of electron motion in the EM field of the cosine form 
$F_0\cos{\Omega t}$. Similar values
for the excursion radius are obtained for all EM fields given 
by \Eref{comp} we consider below. Thus, the electron 
moves predominantly far from the nucleus, and neglect of the 
Coulomb potential in the classical equations of motion is
legitimate.

\begin{figure}[h]
\epsfxsize=10cm
\epsffile{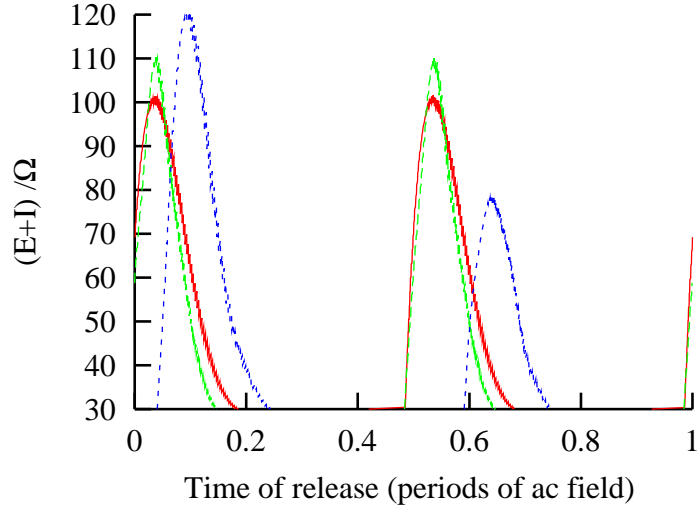}
\caption{(Color online)
Classical dependence of the quantity $ (E+I_{\rm P})/\Omega$
on the time of electron release within an optical cycle
($E$- electron energy at the moment of return to the
nucleus, $I_{\rm P}=0.196$~a.u.- ionization potential of the
Li atom).  The three sets of curves correspond,
respectively, to the pure cosine wave --  solid (red) line; odd
harmonics in Table~\ref{tab1} --  dashed (green) line; odd and
even harmonics in Table~\ref{tab1} --  short (blue) dash.  }
\label{fig1}
\end{figure}

For the pure cosine form of the EM field, the classical procedure
described above
leads to the typical dependence of the kinetic energy at the
moment of return on the time of release shown in
\Fref{fig1} by the solid (red) line. 
For convenience, in \Fref{fig1} we plot not just the kinetic
energy itself, but the quantity $ N=(E+I_P)/\Omega$, which
gives us the order of the harmonic corresponding to given kinetic
energy $E$.
The solid curve in \Fref{fig1} shows that, for the
parameters we chose, the maximum harmonic order
is approximately $N_{\rm cut-off}\approx 100$, which is a
visualization of the well-known $I_{\rm p}+3.17 U_{\rm p}$
cut-off rule.

For the set of parameters in \Eref{comp}, defining the EM
field different from the pure cosine wave, we proceed as
follows. For each set of parameters in  \Eref{comp},
subject to the 
constraint of
the fixed intensity, so that the field intensity had the
same value as in the case of the pure cosine wave,
we can
compute a maximum kinetic energy of the returning electron.
This gives us a function defined on the set of the parameters $a_k$
in  \Eref{comp}.
We look for the
maximum of this function using the gradient ascent method,
giving as a starting values of the independent variables
some particular set of the 
parameters in \Eref{comp}, satisfying the fixed intensity
constraint. 
This procedure is guaranteed to converge to a local maximum. 
Since the convergence is generally quite fast, and requires only
modest computational effort, 
we can repeat the procedure many times with different 
starting values, untill we can be reasonably sure, that 
we have found the global maximum.

We perform two calculations of this kind.  In the first, we
impose an additional restriction that 
only the terms with odd $k$-values
are to be present in \Eref{comp}. This ensures that the
resulting HHG spectrum contains only odd harmonics of the main
frequency.  In the
second calculation, we retain the terms 
with both odd and even $k$-values
in the expansion \eref{comp}. 
In this case, the resulting HHG spectrum
contains even harmonics as well, since symmetry of the Hamiltonian,
which for the case when only odd harmonics are present
in \eref{comp}, leads to only odd harmonics in the HHG spectrum, 
is broken by superimposition of fields of $\Omega$ and $2\Omega$
frequencies \cite{evenh}.

The first calculation
was performed with $K=7$, while in the second we chose
$K=5$. The resulting sets of coefficients $a_k$ for which
the maximum of the highest kinetic energy of the returning
electron is attained, are presented in Table~\ref{tab1}. Also
presented is the set consisting of only $a_1$, which defines
the pure cosine wave for the field parameters considered
above. 

\begin{figure}[h]
\epsfxsize=10cm
\epsffile{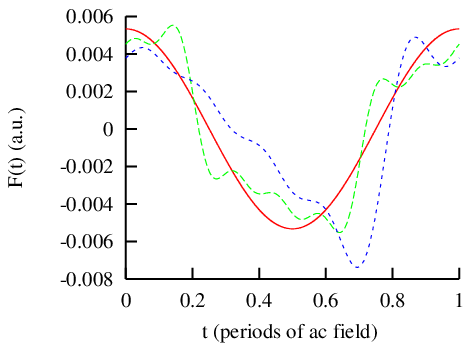}
\caption{(Color online)
EM fields corresponding to the coefficients $a_k$ listed in
Table~\ref{tab1}. The pure cosine wave -- (red) solid line; odd
harmonics in \Eref{comp} with $K=7$ -- (green) dashed line; odd
and even harmonics in
\Eref{comp} with $K=5$ -- (blue) short dash.   }
\label{fig2}
\end{figure}

The degree to which this procedure increases the highest
energy of the returning electron is illustrated in
\Fref{fig1}. Resulting shapes of the driving field $F(t)$,
corresponding to the three cases considered above are
visualized in \Fref{fig2}.

\begin{table}[h]
\begin{tabular} {c crr}
 & cosine wave & odd harmonics & odd and even harmonics \\ 
k & $ a_k\cdot 10^3$ & $  a_{2k-1}\cdot 10^3$ & $ a_{k}\cdot 10^3$ \\
\hline\hline\\
$1$  & 2.665     & $ 2.503-0.076i$ & $2.123-1.033i$  \\
$2$  & 0         & $-0.443-0.566i$ & $0.403+0.754i$ \\
$3$  & 0         & $ 0.061-0.385i$ & $-0.558+0.271i$ \\
$4$  & 0         & $ 0.138-0.264i$ & $-0.302-0.358i$  \\
$5$  & 0         & 0                   &  $0.224-0.248i$  \\
\end{tabular}
\caption{Coefficients in \Eref{comp} for which
the highest kinetic energy of the returning electron is
maximized. The second column: pure cosine
wave; the third column: odd harmonics with $K=7$;
the fourth column: odd and even harmonics with $K=5$ 
\label{tab1}}
\end{table}

As one can see, the set of the parameters corresponding to
only odd harmonics present in \Eref{comp} allows to achieve
a 10\% gain in the position of the cut-off.  The curve
representing dependence of the kinetic energy on the time of
release remains symmetric with respect to the translation
$t\to t+T/2$, as in the case of the pure cosine wave. This
is, in fact, a general property exhibited by the classical
solutions of the equations of motion in the EM field with
only odd harmonics present in \Eref{comp}, which leads to
essentially the same structure of the classical returning
electron trajectories as in the case of the cosine wave.
There are two such trajectories per every half cycle of the
EM field (the so-called "long" and "short" trajectories )
for the plateau region, i.e. for the kinetic energies below
the apex of the corresponding curves in \Fref{fig1}.
There is one trajectory per every half cycle with the
kinetic energy of the returning electron near the apex of
the curves (the cut-off harmonics).

Situation is different for the case of even and odd harmonics present
in  \Eref{comp}. The kinetic energy curves are no longer 
symmetric with respect to the half cycle translation
$t\to t+T/2$. 

One should note, that increase in the cut-off position shows
very little sensitivity to further increase of the number of
terms in \Eref{comp}. If, for example, we used $K=9$ instead
of $K=7$ in the case of only odd harmonics included in
\Eref{comp}, we would have gained additional increase in the
cut-off position of the order of 1\%. Similar observation
applies for the case of even and odd harmonics in
\Eref{comp}.  This indicates, that the low order harmonics
in the series (\ref{comp}) are primarily responsible for the
increase in the cut-off position, and the pulses composed
using the coefficients in Table~\ref{tab1} are optimal
in the sense that no further significant increase in the
cut-off position is possible as long as we rely on the
expansion \eref{comp} for the waveform.

The discussion presented so far was purely classical and
constituted a simple generalization of the 3-step model for the 
case of the EM field given by \Eref{comp}.
Quantum calculation
is needed to confirm the classical results. 
Such calculation is presented in the next section.

\subsection{Quantum calculation}

In this section, we present results of the HHG calculation
for the Li atom for the set of coefficients $a_k$ given in
Table~\ref{tab1}. We use the procedure, which we developed
recently in Ref.~\cite{hhresn} for the solution of TDSE for
realistic atomic targets, which can be described within the
single active electron approximation. For completeness, most
essential features of this procedure are outlined below.

The field-free atom in the ground state is described by
solving a set of self-consistent Hartree-Fock equations
\cite{CCR76}. The field-free Hamiltonian ${\hat H}_{\rm atom}$
in this model is thus a non-local integro-differential
operator. 

The EM field is chosen to be linearly polarized along the
$z$-axis.  We describe the atom-EM field interaction using
the length gauge: $\hat H_{\rm {int}}=zF_z(t)$, where
$F_z(t)=f(t)F(t)$. Function $F(t)$ is given by \Eref{comp}
where we use one of the three 	sets of the coefficients from
Table~\ref{tab1}.  The switching function $f(t)$ smoothly
grows from 0 to 1 on a switching interval $0<t<T_1$, and
is constant for $t>T_1$.  The switching time is $T_1=5T$.

We represent solution of TDSE in the form of an expansion on
a set of the so-called pseudostates:
\begin{equation}
\Psi({\bm r},t)=\sum\limits_{j}
b_j(t) f_j({\bm r}) 
\label{exp}
\end{equation}
This set is obtained by diagonalizing the field-free atomic
Hamiltonian on a suitable square integrable basis
\cite{B94,bstel}:
\begin{equation}
\langle f^N_{i}|{\hat H}_{\rm atom}| f^N_{j}\rangle=E_{i}
\delta_{ij}
\ .
\label{pseud}
\end{equation}
Here the index $j$ comprises the principal $n$ and orbital $l$
quantum numbers, $E_{j}$ is the energy of a
pseudostate and $N$ is the size of the basis.

To construct the set of pseudostates satisfying \Eref{pseud}, we
use either  the Laguerre basis, or the set of 
B-splines (for angular momenta $l>15$), confined to a box of a
size $R_{\rm max}=200$ a.u.
B-splines of the order $k=7$ with the knots located at the
sequence of points lying in $[0,R_{\rm max}]$ are employed.  
All the knots $t_i$
are simple, except for the knots located at the origin and the
outer boundary $R=R_{\rm max}$ of the box. These knots have
multiplicity $k=7$. The simple knots were distributed in $(0,R_{\rm
max})$ according to the rule $t_{i+1}=\alpha t_i+\beta$. The parameter
$\alpha$ was close to 1, so that the resulting distribution of the
knots was almost equidistant.  For each value of the angular momentum
$l$, the first $l+1$ B-splines and the last B-spline resulting from
this sequence of knots were discarded.  Any
B-spline in the set thus decreases at least as fast as $r^{l+1}$ 
and assumes zero value at the outer boundary.

In the present calculation, the system is confined within a
box of a finite size which may lead to appearance of
spurious harmonics in the spectrum due to the reflection of the
wavepackets from the boundaries of the box \cite{hhg1}. One
can minimize this effect by using a mask function or an
absorbing potential.  We use the absorbing potential
$-iW({\bm r})$ which is a smooth function, zero for $r\leq
180$ a.u. and continuously growing to a constant $-iW_0$
with $W_0=2$ a.u. outside this region.

For the EM field parameters which we employed in the
classical treatment of the previous section, the maximum
harmonic order is of
the order of a hundred.  This implies that to describe
accurately formation of all harmonics, we have to retain
pseudostates with correspondingly high angular momenta.  In
the calculation we present below, the pseudostates with
angular momenta $l<120$ were retained in \Eref{exp}.

With the total Hamiltonian and basis set thus defined, the
TDSE can be rewritten as a a system of differential
equations for the coefficients $b_j(t)$ in \ref{exp}. This
system is solved for the time interval $(0,30T)$, where $T$
is a cycle of the EM field, using the Crank-Nicholson method
\cite{crank}.

Finally, the harmonics spectrum is computed as \cite{hhg1}:
\begin{equation}
|d(\omega)|^2=
\left |{1\over t_2-t_1}
\int\limits_{t_1}^{t_2}e^{-i\omega t}d(t)\ dt \right|^2
\ .
\label{hhg}
\end{equation}
Here $\displaystyle d(t)=\langle \Psi(t)|z|\Psi(t)\rangle$
is expectation value of the dipole momentum, limits of
integration $t_1$ and $t_2$ are chosen to be large enough to
minimize the transient effects (we use last 10 cycles of the
pulse duration, i.e., $t_1=20T$, $t_2=30T$).

\section{Results}

\begin{figure}[h]
\epsfxsize=10cm
\epsffile{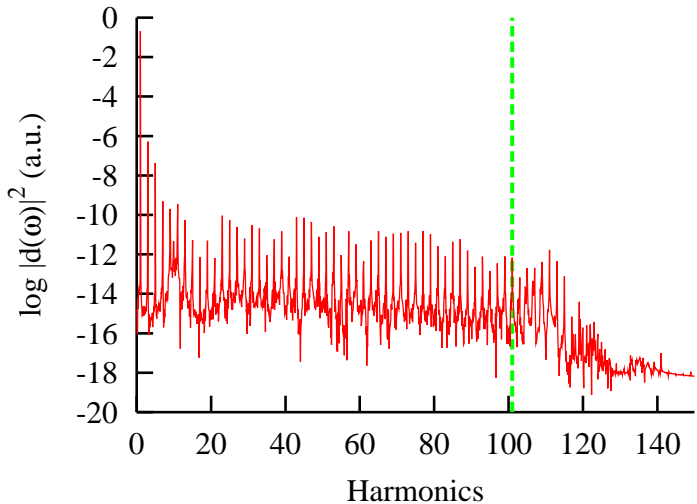}
\caption{(Color online)
Harmonics spectra of Li for the EM fields from Table
\ref{tab1}.  Pure cosine wave ((red) solid line), 
classical cut-off position
marked with the (green) dashed line.}
\label{fig3a}
\end{figure}

\begin{figure}[h]
\epsfxsize=10cm
\epsffile{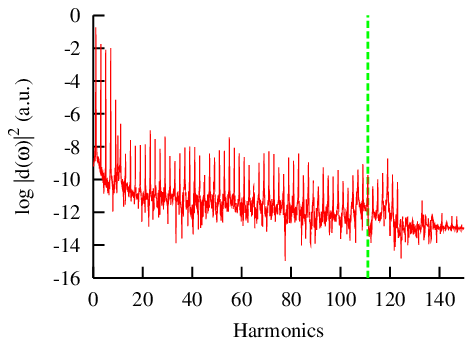}
\caption{(Color online)
Harmonics spectra of Li for the EM fields from Table
\ref{tab1}. 
Odd
harmonics in \Eref{comp} with $K=7$
((red) solid line),
classical cut-off position
marked with the (green) dashed line.}
\label{fig3b}
\end{figure}

\begin{figure}[h]
\epsfxsize=10cm
\epsffile{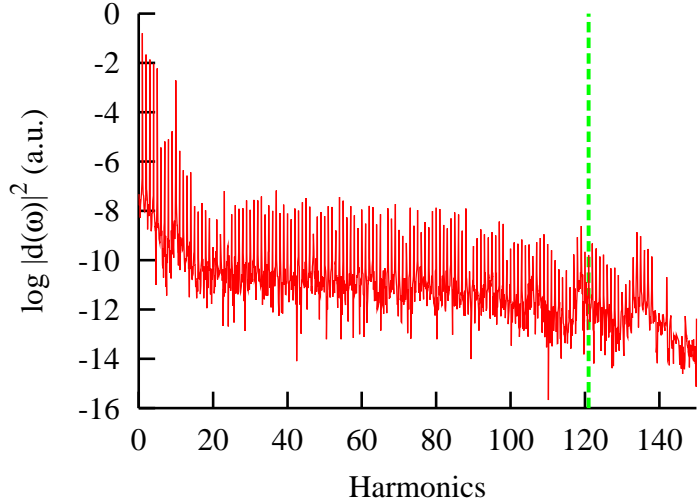}
\caption{(Color online)
Harmonics spectra of Li for the EM fields from Table
\ref{tab1}.
Odd and
harmonics in \Eref{comp} with $K=5$
((red) solid line),
classical cut-off position
marked with the (green) dashed line.}
\label{fig3c}
\end{figure}

In Figures\ {\ref{fig3a},\ref{fig3b},\ref{fig3c}} 
we show the harmonics spectra resulting from
the TDSE calculation for the three choices of the EM field
coefficients listed in Table~\ref{tab1}. We remind the
reader that in all three cases field intensities are equal to
$W=10^{12}$ W/cm$^2$.

\begin{figure}[h]
\epsfxsize=10cm
\epsffile{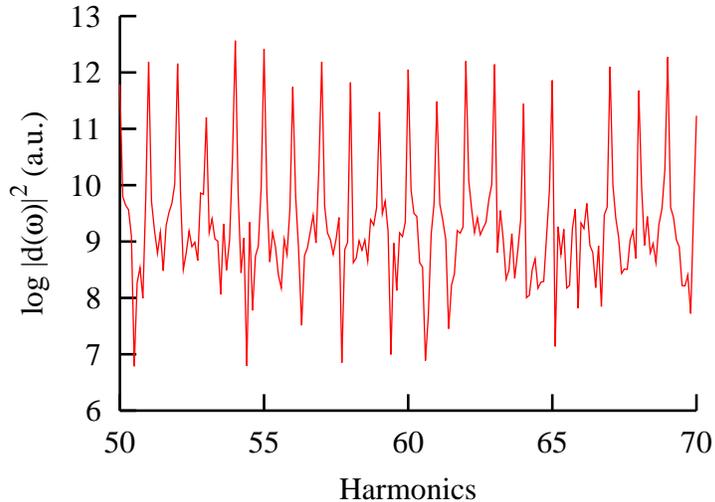}
\caption{(Color online)
Part of the spectrum of Li for
odd and even harmonics in \Eref{comp} with $K=5$.
}
\label{fig4}
\end{figure}

General appearance of these spectra agrees with the
expectations based on the classical results of the previous
section.  One can observe the increase in the cut-off
position for the pulse shaped according to the recipe from
the third column of  Table~\ref{tab1} (only odd harmonics
with $K=7$ in \Eref{comp}), comparing to the cut-off
position for a pure cosine wave of the same intensity.
Cut-off position increases yet further for the pulse
constructed using the set of the coefficients from the
fourth column of Table~\ref{tab1} (odd and even harmonics with
$K=5$ in
\Eref{comp}). In this case, the spectrum contains
harmonics of both odd and even orders. A magnified fragment
of the spectrum illustrating this fact is shown in
\Fref{fig4}.

Quantitatively, the TDSE results for the cut-off positions
are in good agreement with the classical predictions,
summarized in \Fref{fig1}. Use of the pulse constructed
from all harmonics with $K=5$ in \Eref{comp} allows to
increase the cut-off position by about 20\%, in agreement
with the classical analysis given above.

We can, in fact, establish a closer correspondence between
classical and quantum results by performing the
time-frequency analysis of our data.  The techniques used
for this purpose, the wavelet transform
\cite{wavelet4,wavelet3,wavelet2,wavelet1}, or the closely
related Gabor transform \cite{wavelet4,hhres}, offer
possibility to track the process of harmonics formation in
time, combining both the frequency and temporal resolution of
a signal. By using these techniques, we can try to find, in
the quantum domain, the traces left by the classical
trajectories. The fact that such traces may be present,
follows from the quantum-mechanical treatment of the HHG
process given in \cite{hhgd}, where the classical
trajectories naturally appear in the saddle-point
analysis. Such manifestation of the classical trajectories
in the HHG spectra was demonstrated, for example, for the
hydrogen atom \cite{wavelet2}.

We perform our analysis of the HHG process by applying the wavelet
transform of the dipole operator $d(t)$ in
\Eref{hhg}.  This transform 
is defined as \cite{wavelet}
\begin{equation}
T_{\Psi}(\omega,\tau)=\int d(t)\sqrt{|\omega|}
\Psi^*(\omega t-\omega \tau)\ dt\ .
\label{wav1}
\end{equation}
The transform is generated by the Morlet wavelet
$\displaystyle \Psi(x)= x_0^{-1} \exp(-ix) \exp[-x^2/(
2x_0^2)]$.

\begin{figure}[h]
\begin{center}
\begin{tabular}{cc}
\resizebox{50mm}{!}{\epsffile{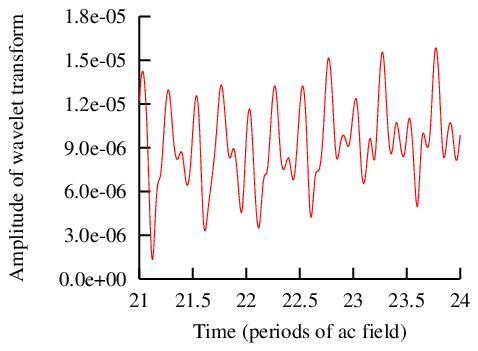}} &
\resizebox{50mm}{!}{\epsffile{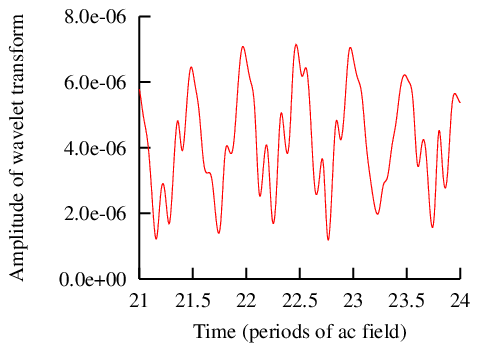}} \\
\end{tabular}
\caption
{(Color online)
Wavelet time-spectrum of Li for the 61-st (left panel) and
101-st (right panel)  harmonics for the pure cosine wave
in \Eref{comp}.
}
\label{fig5}
\end{center}
\end{figure}

\Fref{fig5} presents a well-known picture of the harmonics
formation in time \cite{wavelet2}.  For the plateau
harmonics, the amplitude of the wavelet transform has four
maxima per cycle, corresponding to the two pairs of the
so-called long and short trajectories for the harmonics at
the plateau.  For the near cut-off 101-st harmonic, two
maxima per cycle are present. Those features agree
completely with the classical picture shown in
\Fref{fig1}.

\begin{figure}[h]
\begin{center}
\begin{tabular}{cc}
\resizebox{50mm}{!}{\epsffile{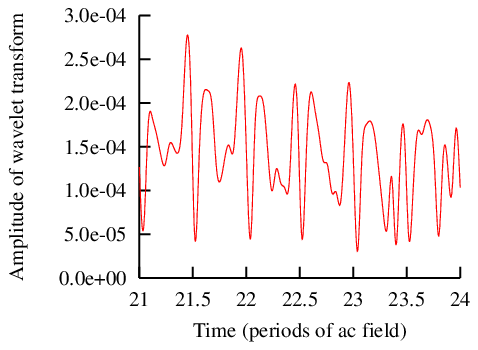}} &
\resizebox{50mm}{!}{\epsffile{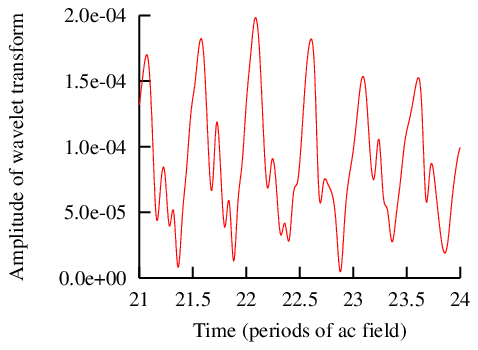}} \\
\end{tabular}
\caption
{(Color online)
Wavelet time-spectrum of Li for the 65-th (left panel) and
99-th (right panel)  harmonics for the pulse
with only odd harmonics in \Eref{comp} ($K=7$).
}
\label{fig6}
\end{center}
\end{figure}

For the pulse containing only
odd harmonics in \Eref{comp}, classical picture of
the dependence of kinetic energy on the time of release, presented
on \Fref{fig1}, is very similar to the curve for the pure
cosine wave. We can expect, therefore, 
results of the wavelet transform in this case to be 
qualitatively similar to those shown  on \Fref{fig5}. 
That this is indeed the
case can be observed from  \Fref{fig6}.

\begin{figure}[h]
\begin{center}
\begin{tabular}{cc}
\resizebox{50mm}{!}{\epsffile{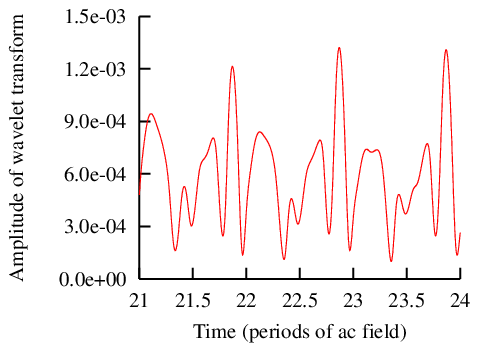}} &
\resizebox{50mm}{!}{\epsffile{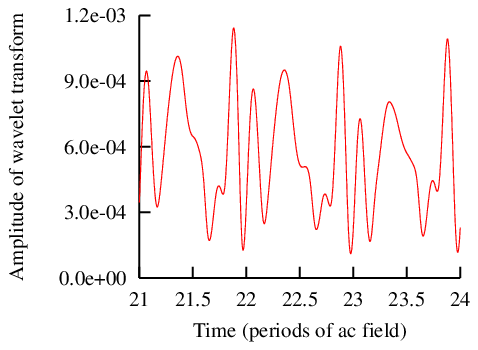}} \\
\resizebox{50mm}{!}{\epsffile{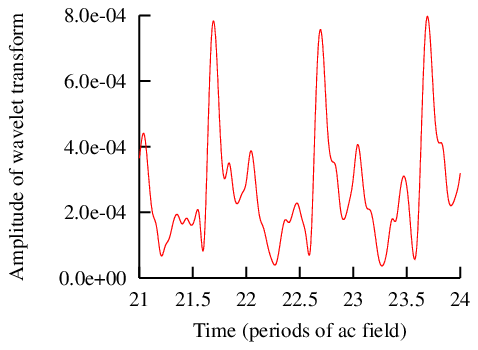}} &
\resizebox{50mm}{!}{\epsffile{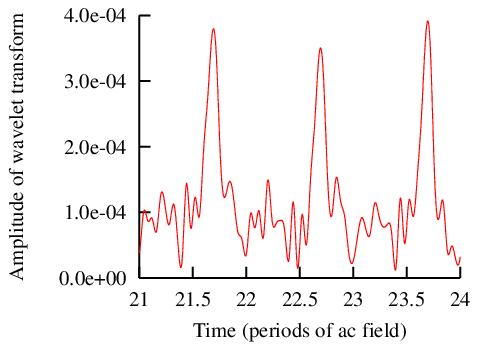}} \\
\end{tabular}
\caption
{(Color online) Wavelet time-spectrum of Li for the 65-th,
75-th, 97-th and 111-th harmonics (from left to right and
top to bottom). The driving field contains terms with both odd and
even $k$-values in \Eref{comp} with $K=5$.  }
\label{fig7}
\end{center}
\end{figure}

For the field waveform containing both odd and even harmonics
in \Eref{comp}, the classical analysis reveals
a somewhat different picture.  As one can see from 
\Fref{fig1}, there are  two pairs of the
classical trajectories per cycle for which kinetic energy of
the returning electron is such, that less than approximately
60 harmonics can be formed. When the harmonic order
increases and reaches the value of
approximately 75 (cut-off region for the smaller maximum of
the corresponding curve in \Fref{fig1}), there are three
returning trajectories per cycle. For higher energies, there
remain only two classical trajectories, which can
participate in the formation of the harmonics.  For higher
yet energy, a single such trajectory exists.

As can be observed from \Fref{fig7}, the quantum calculation
apparently confirms these classical considerations.  Wavelet
spectra do demonstrate that number of maxima per cycle
progressively decreases with the increase of the harmonics
order.

\section{Conclusion}

We demonstrated an increase of the cut-off value for the HHG
process when a superposition of several harmonics of a given
frequency is used to build a waveform of the driving EM
field.  We analyzed the classical returning electron
trajectories for the fields thus constructed.  Such an
analysis shows, that a field spectral composition can be
found, for which  a 20\% increase in the value
of the maximum classical kinetic energy of the recombining
electron is achieved as  compared to the case of a cosine wave
of the same intensity.

TDSE calculation of the HHG spectrum for such a driving
field, performed for the Li atom, confirms the classical
result.  It does demonstrate the increase in the cut-off
value of the order of 20\%.
This value represents a maximum increase which
can be achieved if we restrict the trial waveform to that
given by \Eref{comp} under condition of a fixed
intensity. Indeed, the classical calculation shows, that no
further noticeable increase of the maximum classical kinetic
energy of the recombining electron can be achieved by adding
higher order harmonic terms in expansion
\eref{comp}.

Our result thus presents an upper limit in the increase of the HHG
cutoff achieved for the class of the waveforms given by
\Eref{comp}, i.e. for the waveforms which are periodic with
a given period $T$ and do not contain the DC components.
This suggests, that to achieve more substantial increase
in the HHG cutoff condition, one should use the waveforms which
cannot be described by \Eref{comp}.  Such are the ideal waveform
proposed in \cite{kinsler}, 
for which we should allow the term with $k=0$
to the sum in  \Eref{comp}), or the field configurations containing
subharmonic fields with frequencies $\Omega/2$, as those used in
\cite{hhgmcol2,subharmonic}. As results of these works indicate, a
considerably more important gain in the cutoff energy can be achieved
for such waveforms. These results, and the result obtained in the
present work, allow us to draw the following conclusion.  The strategy
based on the low-frequency (subharmonic) modifications of the waveform
may be more efficient than the strategy relying on introducing
multiple-frequency components in the trial waveform as in
\Eref{comp}. This may provide a useful guide to the problems related
to modification of the high frequency part of the HHG spectrum.

The time-frequency analysis of the results of the TDSE
calculation illustrates the role, which the classical
trajectories play in the formation of the harmonics. The
usual picture of HHG rendered by this technique exhibits
traces of four (for the plateau harmonics) or two (harmonics
near cut-off) trajectories per optical cycle, which
participate in forming a particular harmonic.  In the case
of the waveform constructed from the terms of odd and even order in
\Eref{comp}, the picture revealed by the wavelet analysis is
different.  Number of contributing trajectories in this case
varies with energy in agreement with the classical picture
of \Fref{fig1}. Depending on the harmonics order, there may
be four, three, two or just a single such trajectory.

For a single atom, each 
such a trajectory leads to the formation of a short burst of EM
radiation, producing a pulse train. In the case of the HHG driven by
the single color $T$-periodic EM field, such a train is  a $T/2$ periodic
sequence of bursts, with two bursts on each interval of the length
$T/2$, corresponding to the short and long trajectories within 
a half cycle. For each harmonic order, the contributions of these two 
trajectories interfere strongly, leading to the random distribution
of the phases of different harmonics in the plateau region.

This situation is changed \cite{trains} if propagation
effects are taken into account. Depending on the particular 
propagation geometry, one of the contributions (of either short
or long trajectories) is suppressed, the propagated harmonic
components become locked in phase, and the microscopic signal
is a train with one pulse per every half cycle. 

For the case of 
the waveform with only odd harmonics in  \Eref{comp}, 
propagation should have exactly the same effect, as for the 
single color field. For this waveform, the classic curve in \Fref{fig1}
has exactly the same form as in the single color case, giving rise
to the same set of long and short trajectories per every half cycle
of the laser field. Analysis given in the work \cite{trains} shows, that
propagation effects reduce contribution of one of the  trajectories since 
their phases change differently with laser intensity, and hence 
behave differently in
the nonlinear medium. Depending on the particular geometry, 
contribution of one of the trajectories can thus be reduced. 
The curve in \Fref{fig1} suggests, that in the case of 
the waveform with only odd harmonics in  \Eref{comp} we should
have analogous situation.

For the case of the waveform  with odd and even 
harmonics in  \Eref{comp}, the pulse train produced by the 
single-atom is no longer $T/2$-periodic, but $T$-periodic.
This is clearly seen from  \Fref{fig1}. It is, of course, also
obvious from the fact, that even harmonics are present in the HHG
spectrum in this case, separation of the harmonics is not $2\Omega$
but $\Omega$, consequently
the signal is a $T$-periodic function. On each of the
intervals of length $T$ we have, depending on the number of the
classical trajectories four, three, two, or a single pulse of
different intensities. For the harmonics with orders $N>80$, when,
as seen from  \Fref{fig1}, there are only two trajectories to consider,
propagation should produce essentially the same effect as 
in the case of the single color field. These two trajectories interfere,  
their phases depending differently
on the laser intensity. Thus, as in the single color case, 
propagation may reduce contribution of one of these trajectories, making
harmonics phase-locked.
The microscopic signal will be in this case a train
with one pulse per every cycle.

For the lower order harmonics, when all four trajectories contribute
with different amplitudes and phases, situation is more complicated.
It can hardly be expected, that propagation effects may suppress
contributions of all but one trajectory, and thus eliminate the 
interference of the contributions due to different trajectories
completely.
The harmonics, therefore,
may not be locked in phase in this case.

\section{Acknowledgements}

The authors acknowledge support of the Australian Research Council in
the form of the Discovery grant DP0771312.  Resources of the National
Computational Infrastructure (NCI) Facility were
employed. One of the authors (ASK) wishes to thank the Kavli
Institute for Theoretical Physics for hospitality. 
This work was supported in part by the NSF Grant No.~PHY05-51164


\end{document}